\documentclass[aps,prb,twocolumn,showpacs,superscriptaddress,groupedaddress]{revtex4}
\usepackage{amsmath}
\usepackage{braket}
\usepackage{graphicx}  
\usepackage{dcolumn}   
\usepackage{bm}        
\usepackage{amssymb}   
\usepackage{subfigure}
\begin{document}

\title{Entanglement Entropy on the Cayley Tree}
\author{Yishai Schreiber and Richard Berkovits}
\affiliation{Department of Physics,  Jack and Pearl Resnick Institute,
Bar-Ilan University, Ramat-Gan 52900, Israel}

\begin{abstract}

The properties of the entanglement entropy (EE) of
a clean Cayley tree (CT) are studied.
The EE shows a completely different behaviour
depending on the way the CT is partitioned
into two regions and whether we consider the ground-state or highly
excited many-particle wave function.
The ground-state EE increases logarithmically as function of
number of generation if a single branch is pruned off the tree,
while it grows exponentially if the region around the root
is trimmed. On the other hand, in both cases the highly
excited states' EE grows exponentially.
Implications of these results to general graphs and disordered systems
are shortly discussed.
\end{abstract}

\maketitle

\section{\label{intro}Introduction}

In recent years there has been a renewed interest\cite{Aizenman,DeLuca}
in the problem of Anderson localization on the Cayley tree (CT)
(in the infinite limit known as a Bethe lattice).
This interest is mainly motivated by the connection between
the many body localization phenomenon
\cite{basko06}
and the CT.
Many body localization may be viewed as a
localization problem in
Fock space, where the
coupling between states due to electron-electron interactions resembles the
CT\cite{Altshuler97}.

%

In the context of the Anderson localization on the CT\cite{AbouChacra},
one can envisage that the entanglement entropy (EE) will play
an important role in clarifying the CT localization properties.
In this EE could join other methods such as
level spacing statistics for the CT systems
\cite{DeLuca,sade03,biroli10,biroli12}, and for
many-body interacting systems believed to map on an effective CT in Fock space
\cite{berkovits96,pascaud98,berkovits99,song00,oganesyan07,monthus11,cuevas12}.
Prior to addressing the challenging question of the EE on a disordered
one must clarify the behavior of the EE for a clean CT.
This is our main goal in this paper.

The EE is a measure of the entanglement between two regions, A and B,
of a system which is in some pure state. This measure is
given by the von-Neumann entropy of the reduced density matrix, $\rho_A$,
of region A:
\begin{equation}\label{von_Neumann}
  S_A=-\mathrm{Tr}(\rho_A \ln \rho_A).
\end{equation}
It has been shown that the EE is a very useful measure
to locate and analyze quantum phase transitions
\cite{amico08,eisert09,lehur08,goldstein11,berkovits12,chu13,berkovits14}.
The ground state EE typically scales like the \emph{boundary area} of the
region \cite{amico08,eisert09}. Thus,
for one dimensional systems, one would expect that the EE will be constant.
This is correct for gaped systems, but for
metallic systems there is a logarithmic correction, and
EE grows like $\ln (L_A)$, where $L_A$ is the size of the sub-system $A$.
For insulators the EE should not depend on the region's size. Therefore
for disordered systems the EE can give an indication for the localization
length $\xi$; we expect the EE to grow logarithmically for $\xi\gg L_A$,
and to saturate for $\xi\ll L_A$\cite{berkovits12}.

Applying this picture to the CT is not straight forward.
Unlike the situation for regular $d$ dimensional systems, where
for any regular simply connected area, the boundary area is
proportional to $L_A^{d-1}$, for the CT things are more
complicated. One expects that a region which is connected to the rest
of the graph only at one point (a branch of the tree), will show a
different EE than a region centered at the root of the tree
where the boundary area grows as $\sim C^{L_A}$ (where $C$ is related to the
coordination number of the tree). In this paper we aim to
calculate the EE for a clean CT as a stepping stone
towards investigating the EE in a disordered CT.
We shall pay special attention to the different behavior
for different regions and the physical meaning of these results.


\section{\label{CT}Cayley Tree}

A CT is \cite{Ostilli} a simple connected undirected graph, 
with no closed loops.
A CT of coordination number $Z$ and $N$ generations is described as follows.
There is a root vertex, which we denote $(1)$. The root is linked to $Z$
vertexes belonging to the $2nd$ generation, which we denote
$(2), (3),...,(Z+1)$. Each of those is linked to another $Z-1$ vertexes
on the $3rd$ generation, so that $(2)$ is related to $(Z+2),...,(Z^2+1)$,
and so on. Every vertex is linked to $Z$ others, except those belonging to
the last generation, which are linked to only one vertex each. Unlike
the \emph{Bethe Lattice}, CT is finite, it's boundary being a
non-negligible part of the entire tree: the total number of vertexes
in the tree is $1+Z\frac{(Z-1)^{N-1}-1}{Z-2}$, while the last generation
contains $Z(Z-1)^{N-2}$ vertexes. Therefore, even for $N\to\infty$ the
boundary is important, as is well known in the context of localization
on CT \cite{mirlin94,sade03}.

\begin{figure}
  \centering
  \includegraphics[width=4cm]{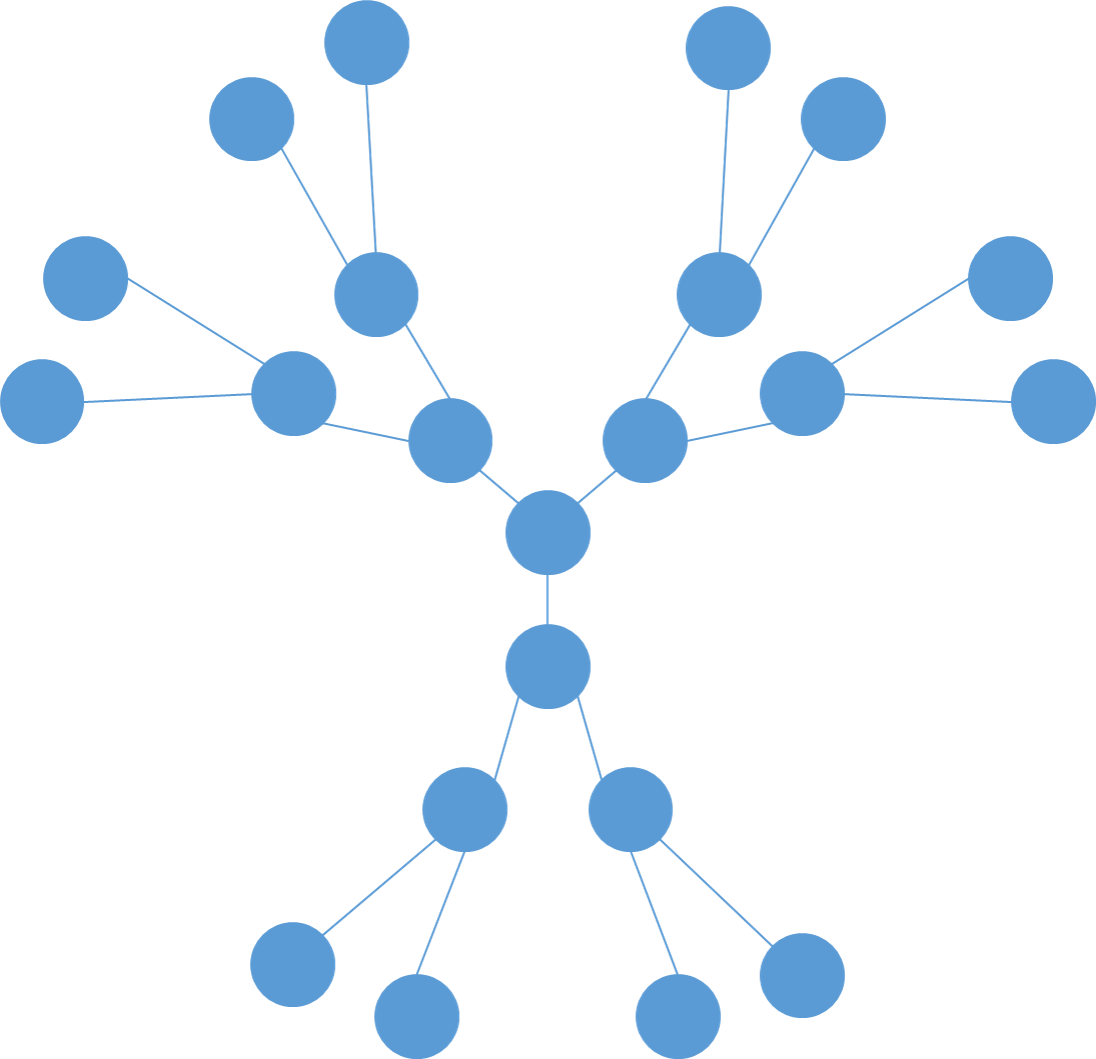}\\
  \caption{A Cayley tree with coordination number $Z=3$ and $N=4$ generations}\label{CT_plot}
\end{figure}

Another noteworthy property of the CT is it's lack of dimension - since
the number of vertexes grows exponentially with $N$, we cannot define the
dimension of this system. On the other hand, any two vertexes are
connected by a single path, what gives the CT a one-dimensional character.
Both of these features are reflected in the scaling of the EE we obtained.

\section {CT Eigenstates}
\label{CT_eigenstates}
First we solve the tight-binding Hamiltonian for a clean CT.
The Hamiltonian can be written, using the notation introduced
previously, in the form
\begin{widetext}
\begin{equation}\label{H_sites}
\hat{H} =-t \sum_{i=2}^{Z+1}(\hat c_1^{\dagger}\hat c_i+h.c.)
  -t  \sum_{g=2}^{N-1}\sum_{i=1}^{Z\left(Z-1\right)^{g-2}} \sum_{j=1}^{Z-1}
( \hat c^{\dagger}_{p_g+i-1} \hat c_{p_{g+1}+(Z-1)(i-1)+(j-1)}+h.c.)
\end{equation}
\end{widetext}
where $\hat c_i^{(\dagger)}$ is an annihilation (creation) operator on
vertex $i$, and $p_g=2+Z\frac{(Z-1)^{g-2}-1}{Z-2}$ is the index of the
first vertex of the $g$'s generation.
It will be convenient to define a new basis, using the operators
$\{\hat{b}^{(\dagger)}\}$, defined as follows. We find an orthonormal
basis in the $2nd$ generation, $\{\hat{b}_2^{2,\nu(\dagger)}\}_{\nu=1}^{Z}$,
and for each such operator we define it's successors in the following
generations $\{\hat{b}_j^{2,\nu(\dagger)}\}_{j=3}^{N}$ as a reflection of
$\hat b_2^{2,\nu(\dagger)}$ in the $j$'s generation. For example, if
$\hat b_2^{2,\nu}=\frac{1}{\sqrt{2}} ( \hat c_2-\hat c_3 )$, then
$\hat b_3^{2,\nu}=\frac{1}{\sqrt{2(Z-1)}}
(\sum_{i=Z+2}^{2Z}\hat c_i-\sum_{j=2Z+1}^{3Z-1}\hat c_j)$ \emph{etc}.
Similarly we find a basis in every generation, composed of operators
other than the successors of the previous generations, and define
its successors in the next generations. We denote each such
operator by $\hat b_j^{\Gamma,\nu (\dagger)}$, where $j$ is the generation
in which this operator acts, $\Gamma$ is the generation in which it's
first fore-father acts, and $\nu$ is it's fore-father's
index inside the $\Gamma$'s generation.
Among these operators there is one important group that originates
from the root, which we shall denote $\hat b_j^{symmetric}$ or $\hat b_j^s$,
$(j=1,...,N)$. They are symmetric in the sense that they treat each
generation as one unit, $i.e.$
\begin{equation}\label{sym_states}
  \hat b_j^s=\left\{
  \begin{array}{rl}
  \hat c_1 & \mbox{for }  j=1\\
  \frac{1}{\sqrt{Z(Z-1)^{j-2}}}\sum_{i=p_j}^{p_{j+1}-1}\hat c_i  &\mbox{for } N\geq j>1
  \end{array}\right. .
\end{equation}
Using this representation, the Hamiltonian becomes
\begin{multline}\label{H_gen}
\hat H=\hat H_s-t\sqrt{Z-1} \left[\sum_{j=2}^{N-1}\sum_{\nu=1}^{Z-1}
(\hat b_j^{2,\nu \dagger}\hat b_{j+1}^{2,\nu}+h.c.)+\right. \\
\left. +\sum_{j=3}^{N-1}\sum_{\Gamma=3}^j \sum_{\nu=1}^{Z(Z-2)(Z-1)^{\Gamma-3}}
(\hat b_j^{\Gamma,\nu \dagger}\hat b_{j+1}^{\Gamma,\nu}+h.c.)\right]
\end{multline}
where the symmetric term
\begin{equation}\label{H_sym}
\hat H_s=-t\sqrt Z(\hat b_1^{s\dagger} \hat b_2^s+h.c.)
-t\sqrt {Z-1} \sum_{j=2}^{N-1}(\hat b_j^{s\dagger} \hat b_{j+1}^s+h.c.).
\end{equation}
Thus all terms, except for $\hat H_s$, are a sum of independent
one-dimensional lattices. One should note the exponentially
increasing number of such lattices. The solution is given by the transformation
\begin{equation}\label{d_states}
  \hat d_k^{\Gamma,\nu}=\sqrt{\frac{2}{N-\Gamma+2}}\sum_{j=\Gamma}^N \sin\left[\frac{(j-\Gamma+1)k\pi}{N-\Gamma+2}\right]\hat b_j^{\Gamma,\nu}
\end{equation}
so that
  \begin{multline}\label{H_final}
    \hat H=\hat H_s- \\
    -t\sqrt{Z-1}\sum_{\Gamma=2}^{N-1}\sum_{\nu}\sum_{k=1}^{N-\Gamma+1} 2\cos\left(\frac{k\pi}{N-\Gamma+2} \right)\hat d_k^{\Gamma,\nu\dagger} \hat d_k^{\Gamma,\nu}
  \end{multline}
 and the spectrum is
 \begin{equation}\label{Ek}
   \varepsilon_{k,\Gamma}=-2t\sqrt{Z-1}\cos\left(\frac{k\pi}{N-\Gamma+2} \right).
 \end{equation}
 For the symmetric term (\ref{H_sym}) a similar solution was obtained
by Chen \emph{et al.}\cite{Chen}, including the boundary condition
for finite $N$.

\section{\label{EE}Entanglement Entropy}

\begin{figure}
  \centering
  \includegraphics[width=5cm]{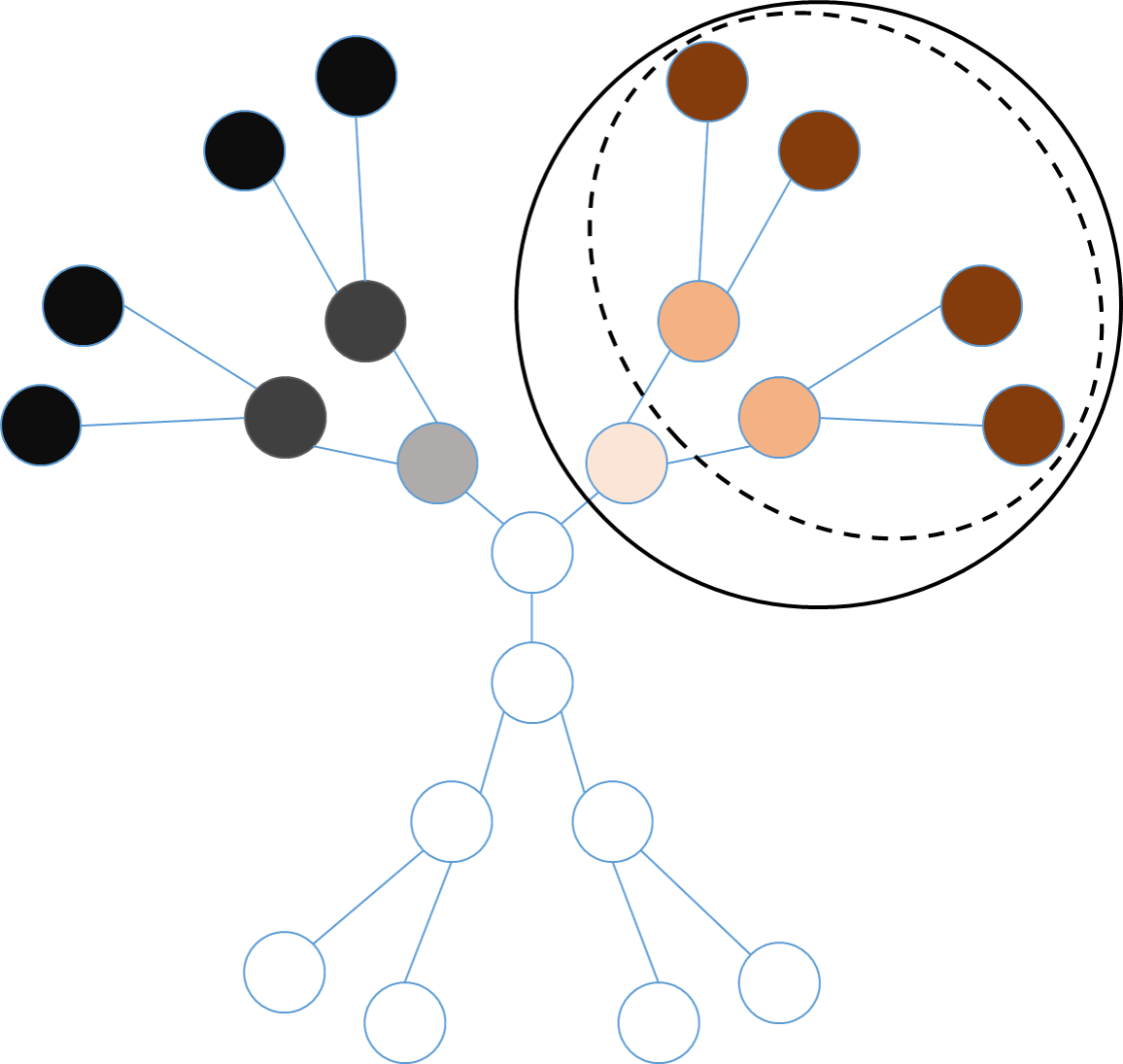}\\
  \caption{A typical eigenstate in a CT with $Z=3$, when we take one branch (solid line) or a part of it (dashed) to be the subsystem. (Color online)}\label{CT_AS_1_branch}
\end{figure}

In order to calculate the EE we use the relation \cite{Latorre}
\begin{equation}\label{EE_cor}
 S=\sum_l -\lambda_l\ln \lambda_l-(1-\lambda_l)\ln (1-\lambda_l)
\end{equation}
where $\{\lambda_l\}$ are the eigenvalues of the correlation matrix
\emph{in region A}. For a one-particle state, the correlation can
be written in the form $\hat C=\ket{\psi_L}\bra{\psi_L}$, where
$\ket{\psi_L}$ is the eigenstate in region A. Thus one eigenvector
is $\ket{\psi_L}$, with corresponding eigenvalue
$\lambda=\braket{\psi_L | \psi_L}$. Since $\hat C$ is Hermitian, any
other eigenvector is orthogonal to $\ket{\psi_L}$, and its corresponding
eigenvalue therefore vanishes, so that the EE is simply
$S= -\lambda\ln \lambda-(1-\lambda)\ln (1-\lambda)$.
We have calculated the EE for several representative
one-particle states in a CT of $Z=3$. A typical state is:
 \begin{multline}
   \ket{\psi_1^{2,1}}=\hat d_1^{2,1\dagger}\ket{\emptyset}= \\
   =\sum_{j=2}^N \frac{1}{\sqrt{2^{j-2}N}}\sin\left[\frac{(j-1)\pi}{N}\right]\cdot\\
 \cdot\left[\sum_{i=p_j}^{p_j+2^{j-2}-1}\hat c_i^{\dagger}-
 \sum_{l=p_j+2^{j-2}}^{p_j+2^{j-1}-1}\hat c_l^{\dagger}\right]\ket{\emptyset}
 \end{multline}
which resides on two of the branches. For this state, we let one
branch be region A (see Fig. \ref{CT_AS_1_branch}), and obtain
$\lambda=\braket{\psi_L | \psi_L}=\frac{1}{2}$,
and $S=-\ln\frac{1}{2}=\ln2$, regardless of the number of generations.
Another possible partition is to define region A as a part of one branch (see in Fig. \ref{CT_AS_1_branch}). The
eigenvalue $\lambda_M$ for this partition where region A is defined
as the last $M$ generations
of one branch is
\begin{equation*}
  \lambda_M=\frac{M}{2N}+\frac{1}{4N}\left(1-\frac{\sin\left[\frac{\pi}{N}\left(2M+1\right)\right]}{\sin\left(\frac{\pi}{N}\right)}\right),
\end{equation*}
which, in the appropriate limit, is proportional to the relative number
of generations in region A.

\begin{figure}
  \centering
  \includegraphics[width=7cm]{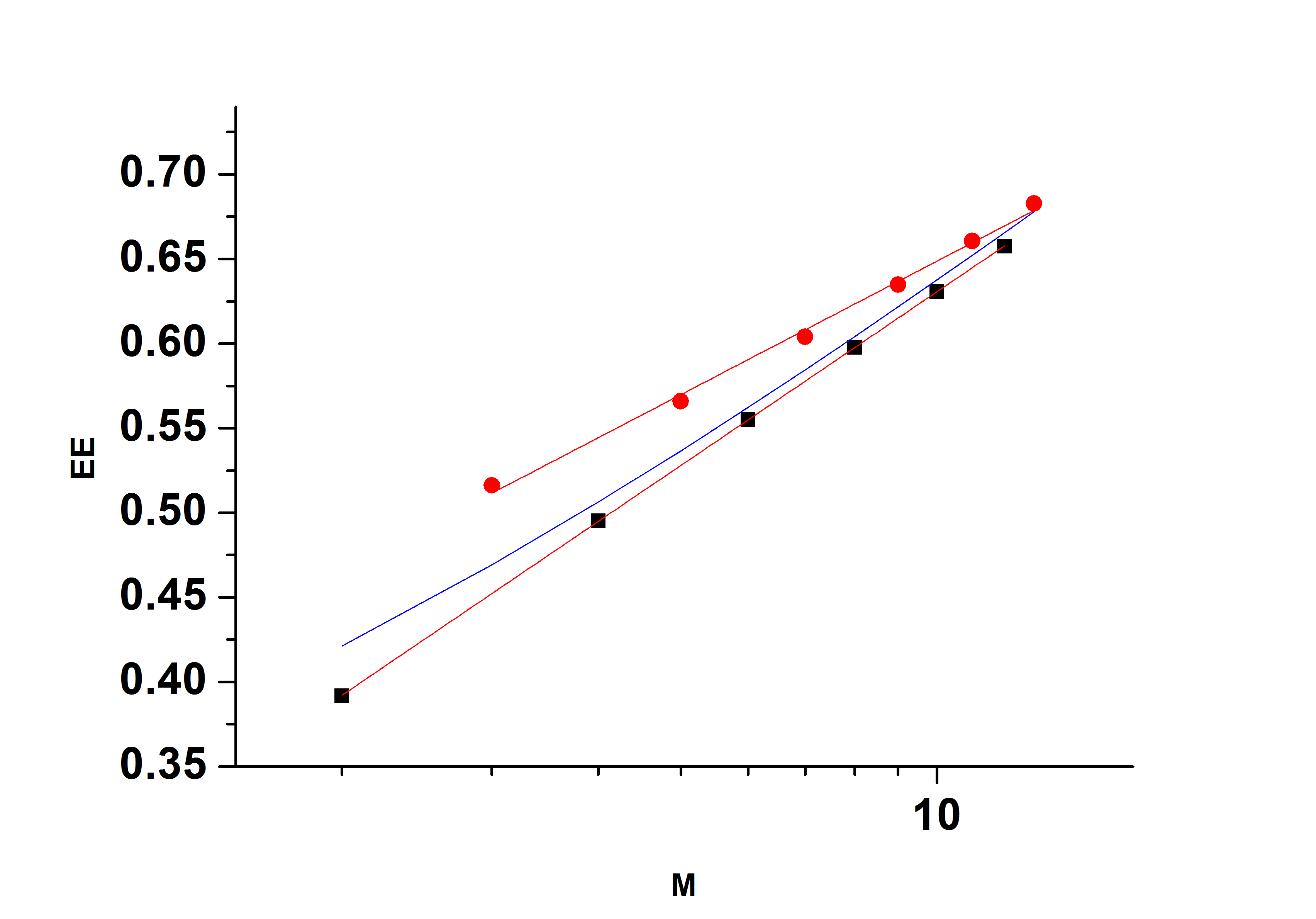}\\
  \caption{Ground state EE between one branch and the rest for $Z=3$ as
function of $M$ the number of generations in the subsystem. H
ere $M=N-1$. The black squares
represent an even
number of generations and the red circles represent
odd $M$'s, separated in order
to emphasize the alternating term. The red lines correspond to
$0.15\ln\left(M\right)+0.29$ for even $M$'s and
$0.11\ln\left(M\right)+0.39$ for odd $M$'s. The blue line
in between is $\frac{1}{6}\ln\left(N\right)+0.24$, which corresponds
to the asymptotic behavior of the EE.
}\label{EE_1_branch}
\end{figure}

\begin{figure}
  \centering
  \includegraphics[width=5cm]{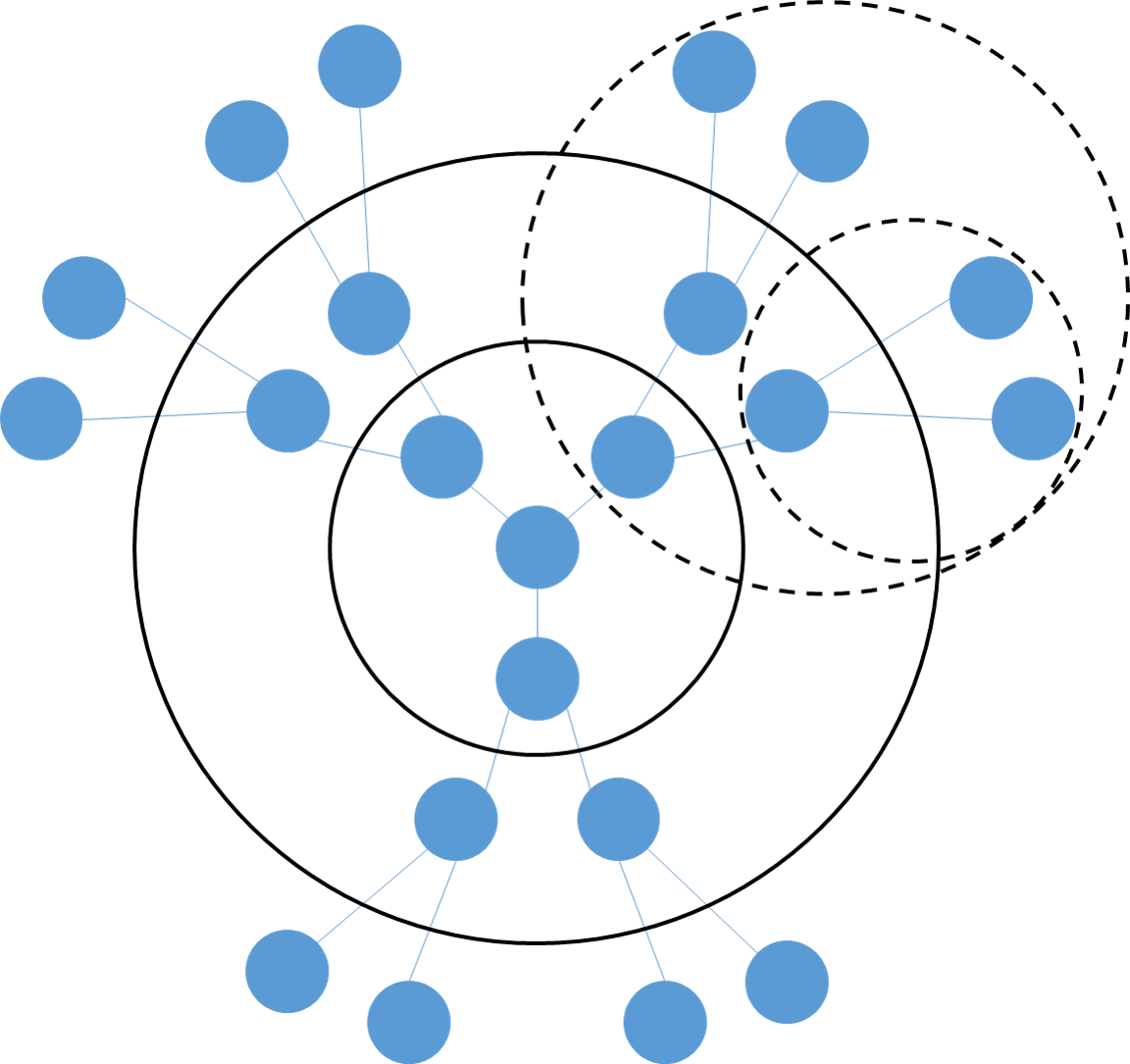}\\
  \caption{Different ways to partition the CT: taking $M$ generations around
the root to be region A (solid circles) or taking a
branch (big dashed circle) or a sub-branch (small dashed circle).}
  \label{BranchingPic}
\end{figure}

For the more challenging case of a many-particle state, the correlation
matrix becomes more complicated, so we diagonalize it numerically.
The many particle state can be written as
\begin{equation}
  \ket{\psi}=\prod_{\Gamma,\nu,k}\hat d_k^{\Gamma,\nu\dagger}\ket{\emptyset},
\end{equation}
where $\Gamma,\nu,k$ are the indicies of the occupied single particle states,
leading to a correlation matrix
\begin{equation}
C_{n,m}=\braket{\hat{c}_n^\dagger \hat{c}_m}= \bra{\emptyset}\prod_{\Gamma,\nu,k}\hat d_k^{\Gamma,\nu} \hat{c}_n^\dagger \hat{c}_m \prod_{\Gamma',\nu',k'}\hat d_{k'}^{\Gamma',\nu'\dagger}\ket{\emptyset}.
\end{equation}
Let us
first consider the ground state EE of a single branch of the CT
(see in Fig. \ref{BranchingPic}) for
the half-filled case.
Since a branch is connected
to the rest of the system only at one point,
one would expect from the area law that the EE will be constant.
Nevertheless, since the CT has no gap, which is similar
to the situation in metallic 1D system, a logarithmic correction
is expected. Indeed,
we find a logarithmic dependence for the CT, but unlike the 1D system
it does not depend on the size of the system, but rather on
the number of generations $N$ in the CT (or, equivalently, on the
number of generations $M$ in the subsystem, where in this case $M=N-1$).
Specifically, for $Z=3$,
$S_A\sim \frac{1}{6}\ln N +\left( -1\right)^N \gamma\left(N\right)$,
where $\left( -1\right)^N\gamma\left(N\right)$ is a decaying alternating
term typical to systems with open boundary conditions\cite{Laflor}
(Fig. \ref{EE_1_branch}).
The pre-factor $\frac{1}{6}$ of the
logarithmic term is reminiscence of the behavior of an open boundary 1D
system, where if one bisects a system of length $L$ in the middle
the EE grows as  $\frac{1}{6}\ln L$ \cite{amico08,eisert09}.
The fact that for the EE on a CT the number of generations $N$ plays the
same role as the length $L$ for a 1D system stems from the structure
of the eigenstates of the CT demonstrated
in Sec. \ref {CT_eigenstates}.
It is interesting to point out that for the CT the
ratio of the sites on a branch to
the overall number of sites is nearly constant ($\sim 1/Z$)
for any generation $N$ larger than $O(1)$, again similar
to the situation for the bisected 1D system.

\begin{figure}
  \centering
  \includegraphics[width=7cm]{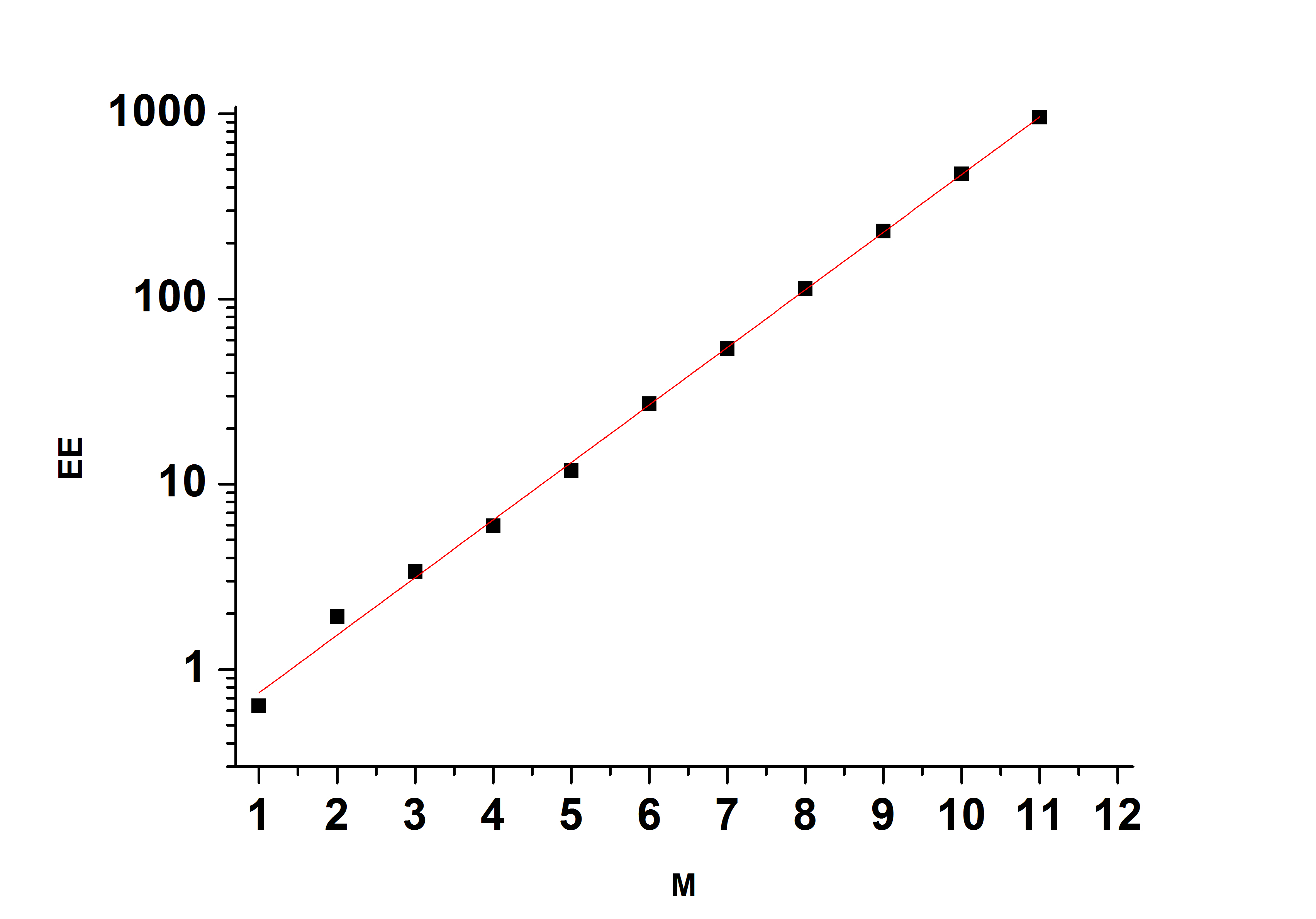}\\
\caption{EE between one branch and the rest of the CT with $Z=3$
for a highly excited state represented by the black squares.
The EE grows as $0.39\exp\left(0.71M\right)-0.65\approx 0.39\times 2^{1.02M}-0.65$ (red line) since for an excited state the EE is proportional to the volume.}
\label{EE_1branch_excited}
\end{figure}

On the other hand, for an excited state we find an exponential growth
of the EE (Fig. \ref{EE_1branch_excited}), reflecting the growth
in the number of sites per generation. This is expected since for an
excited state the EE should obey a volume law. Nevertheless,
this is in stark difference to the one dimensional case for which excited
states EE grows linearly. Thus, for the EE of a branch of the CT
there is a huge change between the ground state EE and the excited state
EE not seen in more standard models. This may lead to a huge influence
of localization on low-lying excitations \cite{berkovits14}.

%
\begin{figure}
  \centering
  \includegraphics[width=7cm]{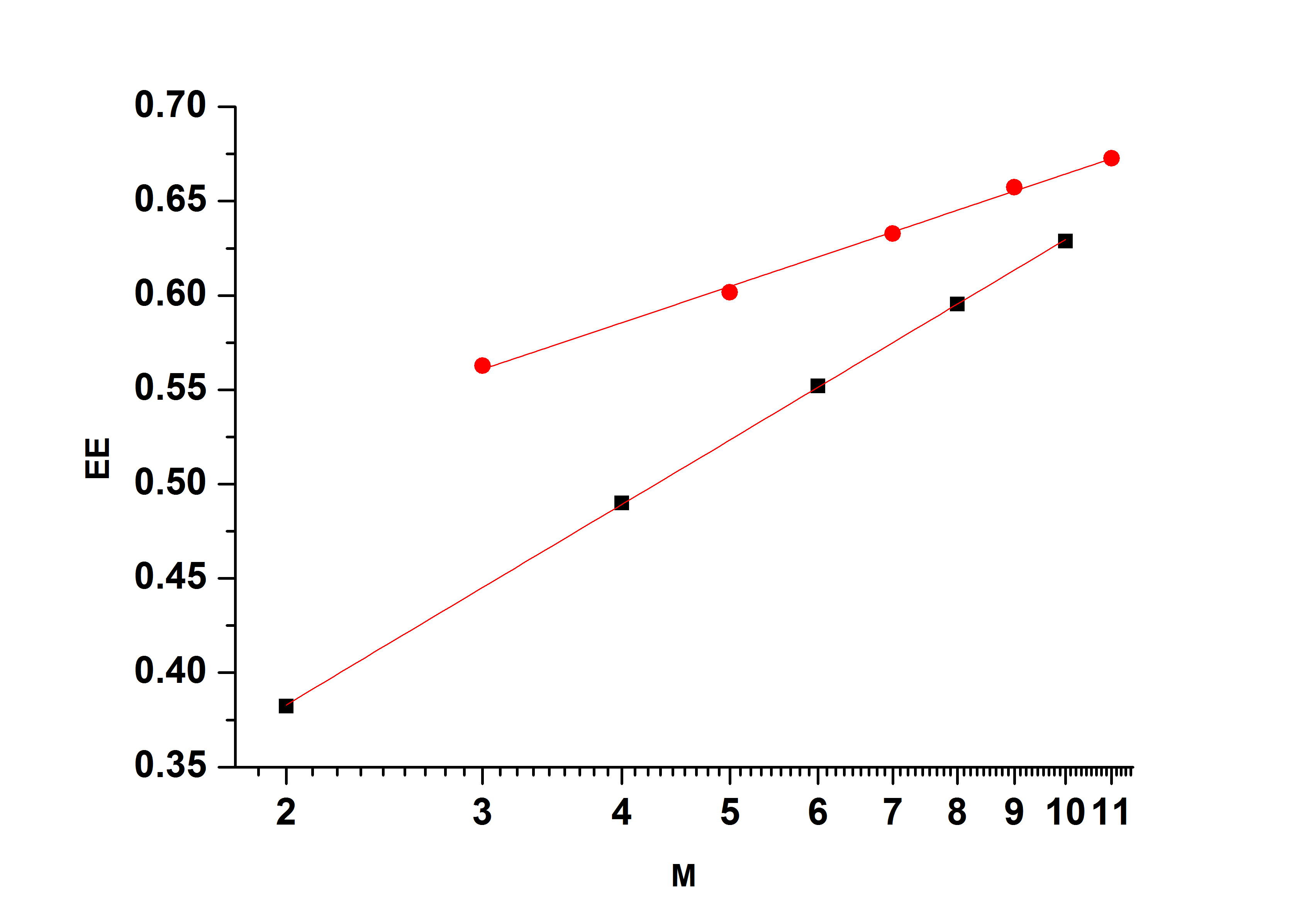}\\
\caption{EE when we cut the CT to sub-branches with $M$ generations,
where the total number of generations in the tree is $N=13$.
The black squares represent even $M$'s, with the red line corresponding
to $0.15\ln\left(M\right)+0.28$, and the red circles are for
odd $M$'s, with the line corresponding to $0.086\ln\left(M\right)+0.47$.}
\label{branching}
\end{figure}

A similar form of dissection is to define a site residing $M$ generations
away from the boundary and all the sites connecting it to the boundary
as region A, so that A is connected to the rest of the tree only at
one point (see in Fig. \ref{BranchingPic}), just like in the previous case.
As expected, we get the logarithmic dependence for this case
also (Fig. \ref{branching}).

\begin{figure}
  \centering
  \includegraphics[width=7cm]{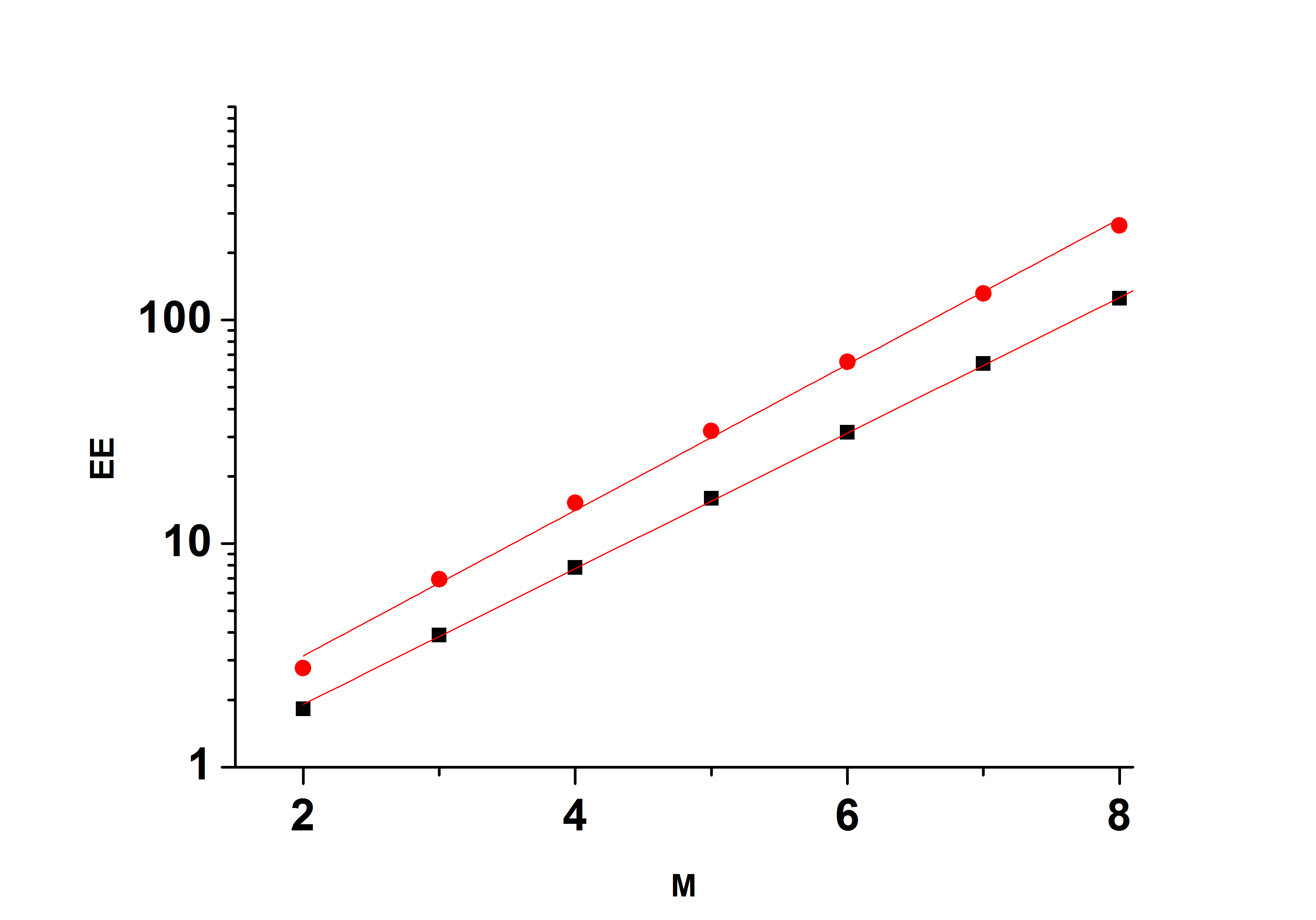}\\
\caption{EE between the first $M$ generations around the root
(including the root itself)
and the rest of the CT with $Z=3$
for the ground state (black squares) and for a highly
excited state (red circles).
In both cases the EE grows exponentially
(the lines, corresponding to $0.55\exp\left(0.68M\right)-0.44 \approx
0.55\times 2^{0.98M}-0.44$ and $1.04\exp\left(0.69M\right)-1.39\approx
1.04\times 2^{M}-1.39$, respectively), since there is no qualitative
difference in the growth as function of $M$ between the interface area and
the volume.}
\label{EE_generation}
\end{figure}

A different obvious way to cut the CT into two different regions is
to define the root and the sites belonging to the first $M$ generations
as region A,
and the rest of the sites belonging to the higher generations as region $B$ (Fig. \ref{BranchingPic}).
In this case the area of contact between the two regions is proportional
to $Z(Z-1)^{M-2}$ for $M\geq2$, and therefore according to the area law
we expect the
ground state EE to be exponentially dependent on $M$. Indeed,
in this case the EE behaves as $S_A\sim \exp{\left(\alpha M\right)}$
(see Fig. \ref{EE_generation}).
Since for such a dissection the volume (number of sites) within region
A is equal to $1+Z\frac{(Z-1)^{M-1}-1}{Z-2} \sim (Z-1)^{M-1}$, i.e.,
grows exponentially with $M$. Also for a highly excited state we find
the EE to follow  $S_A\sim \exp{\left(\alpha^{\prime} M\right)}$, as expected.
Thus, for this form of dissection there is
no qualitative difference between the EE of the ground state
and the behavior of an excited state as function of $M$.

\section{Discussion}

In this paper we
calculated the EE for the ground state and excited states of
a clean CT. Utilizing an exact expression for the single-particle
eigenfunctions of the CT developed in this paper in order to
to obtain a correlation matrix from which the EE
of a many-particle state can be calculated,
it has been demonstrated that the EE may show dramatic differences
depending on the way the CT is partitioned
into two regions. If the CT is dissected
into a branch connected in a single point to the CT, a logarithmic dependence
on the number of generations in the branch is found, which reflects the
one dimensional character of the states of the CT. 
On the other hand, excited states on the
branch show a completely different EE behavior,
namely an exponential dependence on the number of generations
in the branch. This logarithmic to exponential change in the entanglement
between the ground state and the excited state is unique to the entanglement
on the CT.
On the other hand,
the ground state EE shows a completely different behavior for a different
sort of partition - where the root and the first $M$ generations are considered
as one region. In this case
an exponential dependence on $M$ is found, which reflects the growth
in the number of sites on the surface. In contrast to dissecting by branch,
here the excited states EE (which remains the volume law) does not show a
dramatic difference when compared to the ground state EE.

Thus, how one defines the separation into two regions has a crucial
influence on the behavior of the EE.
These large fluctuations in the EE could become even stronger
for other types of complex (random) networks, such as small world graphs,
Erd\"os-R\'enyi graphs and scale free networks \cite{albert02}.

These behaviors predict a strong influence of disorder on the EE of
a CT. It has been shown \cite{berkovits12} that the localization length
in 1D may be inferred from the saturation of the EE on the length scale
of the localization length. For the CT one expects that for a
region around the root and the first $M$ generations where the increase
in the EE is exponential this tool will be more sensitive than
for the 1D system where the growth is logarithmic. It will be also very
interesting to compare the effect of the localization on the EE of a branch.
Will it be similar to the 1D case? Another question is whether the
non ergodic phase predicted for the CT \cite{DeLuca} may be detected
in the behavior of the EE.

\begin{acknowledgments}
Financial support from the Israel Science Foundation (Grant 686/10) 
is gratefully acknowledged.
\end{acknowledgments}

\end{document}